# Preamble Design for Joint Frame Synchronization, Frequency Offset Estimation, and Channel Estimation in Upstream Burst-mode Detection of Coherent PONs

Yongxin Sun, Hexun Jiang, Yicheng Xu, Mengfan Fu, Yixiao Zhu, Lilin Yi, Weisheng Hu, and Qunbi Zhuge, *Senior Member*, *IEEE*, *Senior Member*, *Optica*

*Abstract*—Coherent optics has demonstrated significant potential as a viable solution for achieving 100 Gb/s and higher speeds in single-wavelength passive optical networks (PON). However, upstream burst-mode coherent detection is a major challenge when adopting coherent optics in access networks. To accelerate digital signal processing (DSP) convergence with a minimal preamble length, we propose a novel burst-mode preamble design based on a constant amplitude zero auto-correlation sequence. This design facilitates comprehensive estimation of linear channel effects in the frequency domain, including polarization state rotation, differential group delay, chromatic dispersion, and polarization-dependent loss, providing overall system response information for channel equalization pre-convergence. Additionally, this preamble utilizes the same training unit to jointly achieve three key DSP functions: frame synchronization, frequency offset estimation, and channel estimation. This integration contributes to a significant reduction in the preamble length. The feasibility of the proposed preamble with a length of 272 symbols and corresponding DSP was experimentally verified in a 15 Gbaud coherent system using dual-polarization 16 quadrature amplitude modulation. The experimental results based on this scheme showed a superior performance of the convergence acceleration.

## I. INTRODUCTION

T he optical access network has been continuously evolving, driven by advanced services and applications, such as 5G/6G mobile x-haul, 4K/8K video streaming, and cloud computing [1], [2], [3]. As a cost-effective point-to-multipoint technology, passive optical network (PON) has been proven to be a successful solution for broadband network connections between service providers and users. To meet the growing broadband demands of optical access networks, it is foreseeable that 100 Gb/s/λ and beyond PON will be required [4], [5]. Currently, 50G PON based on intensity modulation and direct detection (IMDD) is progressing toward commercial deployment [6], [7]. However, when PON evolves to 100G and beyond, IMDD systems face challenges in meeting power budget requirements [8]. As coherent detection significantly improves receiver sensitivity, the coherent PON is therefore considered as a promising candidate for future PONs [9], [10].

In addition to significantly improving receiver sensitivity, coherent detection also allows the use of more flexible digital subcarrier multiplexing technologies [11], [12], [13], enabling frequency-division multiple access (FDMA) with flexible bandwidth allocation and notably low latency [14], [15], [16]. Today, PONs have been deployed worldwide to facilitate various fiber-to-the-home (FTTH), fiber-to-the-room (FTTR), and fiber-to-the-building (FTTB) services, predominantly based on time-division multiple access (TDMA) [17]. The TDMA PON accommodates a large number of access points by allocating time slots. However, this scalability comes at the cost of increased latency and the need to deploy high-bandwidth transceivers. To solve these problems, time-and-frequency division multiple access (TFDMA) has been proposed to synthesize the inherent advantages of TDMA and FDMA. Therefore, TFDMA PON can effectively meet the diverse latency and bandwidth requirements, providing a platform for highly flexible network services [18], [19].

Although coherent systems offer numerous advantages in next-generation PON, several significant challenges need to be addressed. Upstream burst-mode detection is one of the key challenges in coherent TDMA PONs [8], [10]. In a downstream situation, signals are generated and modulated in the optical line terminal (OLT) and then continuously broadcast to all users over the fiber network. Receivers in different optical network units (ONUs) detect and demodulate signals in their time slots separately [20]. In a TDMA PON upstream situation, the OLT is tasked with receiving signals from different ONUs, burst by burst. Typically, these burst signals have different optical powers, laser frequency offsets (FO), states of polarization

Manuscript received XX, XX, XXXX; revised XX, XX, XXXX; accepted XX, XX, XXXX. Date of publication XX, XX, XXXX; date of current version XX, XX, XXXX. This work was supported in part by the National Key R&D Program of China under Grant 2022YFB2903500, in part by Shanghai Pilot Program for Basic Research - Shanghai Jiao Tong University under Grant 21TQ1400213, and in part by the National Natural Science Foundation of China under Grant 62175145. *(Corresponding author: Qunbi Zhuge.)*

The authors are with the State Key Laboratory of Advanced Optical Communication Systems and Networks, Department of Electronic Engineering, Shanghai Jiao Tong University, Shanghai 200240, China (e-mail: sunyongxin@sjtu.edu.cn; jianghexun@sjtu.edu.cn; yicheng.xu@sjtu.edu.cn; mengfan.fu@sjtu.edu.cn; yixiaozhu@sjtu.edu.cn; lilinyi@sjtu.edu.cn; wsh@sjtu.edu.cn; qunbi.zhuge@sjtu.edu.cn).

(SOP), and clock frequencies [10]. To ensure the efficiency and reliability of detecting each burst frame, digital signal processing (DSP) must converge quickly in the allocated time slots. However, due to the excessive convergence time, conventional DSP in long-reach point-to-point links is no longer suitable for coherent burst-mode reception. To accelerate the DSP convergence, using an effective, reliable, but short preamble with data-aided low-complexity DSP is considered a practical approach.

Several specially designed preambles have been proposed to help accelerate the DSP convergence [21], [22], [23]. For example, in [21], coherent upstream burst-mode detection was demonstrated in a 100 Gb/s polarization-division-multiplexed-quadrature-phase-shift-keying (PDM-QPSK) TDMA PON. The preamble length was reduced to 1792 symbols by sharing the training sequences (TS) for multiple DSP sections. As for TFDMA PON, an overall burst frame structure with a 416-symbol preamble has been proposed in [22] to achieve fast DSP convergence. In [23], time-domain subcarrier demultiplexing and channel estimation (CE) were realized in TFDMA PON, achieving fast DSP convergence with a 496-symbol preamble. However, in most previous works, the general channel estimation only considers a simple SOP matrix in the time domain, so under some extreme conditions with large differential group delays (DGD), the equalizer trained by least mean square (LMS) algorithm still requires a long TS for convergence. In addition, the traditional TS used for SOP estimation requires that only one training block is transmitted in one polarization, while no signal is present in the other polarization, which can be expressed as $[TS_X, 0; 0, TS_Y]$. As the power envelope varies for certain intervals of time, this scheme is suboptimal for the modulator and induces more nonlinear interference due to the power fluctuation [24]. In the International Telecommunication Union-Telecommunication Standardization Sector (ITU-T) standard of 50G-PON, the burst-mode overhead time has been defined on the timescale of 100 ns [25]. Considering that the preamble time is equal to the length divided by the symbol rate, for low symbol rate subcarriers in TFDMA PON, a further reduction of the timescale length is necessary.

In this paper, we propose an efficient and effective preamble based on constant amplitude zero auto-correlation (CAZAC) sequences to address these issues. In contrast to previous works, this scheme provides two distinct advantages. First, this preamble supports overall channel estimation in the frequency domain (FD). This means besides the rotation of SOP (RSOP), many other linear optical effects, such as polarization mode dispersion (PMD), chromatic dispersion (CD), and polarization-dependent loss (PDL) can be estimated simultaneously. This comprehensive CE helps to obtain much more accurate initial coefficients for the FD equalizer. In this way, the length of training sequences used in equalization convergence can be significantly reduced. Second, this preamble structure is specifically designed to share the same training unit for three DSP functions: frame synchronization (FS), frequency offset estimation (FOE), and channel estimation. This results in a further reduction of the whole preamble length.

This paper is an extension of our prior report on OFC 2024 [26]. Herein, we introduce more details about the preamble properties and the principles of the key DSP. We also present the process of exploring the optimal preamble length configuration and add some additional simulation results under different channel conditions. To validate our scheme, we implemented burst-mode detection for a 15 Gbaud dual-polarization 16 quadrature amplitude modulation (DP-16QAM) upstream signal, utilizing a 272-symbol preamble with a duration of 18.13 ns (272 symbols/15 Gbaud).

## II. PRINCIPLES

In this section, we introduce the detailed preamble design and corresponding DSP at the receiver side. As shown in Fig. 1(a), the overall structure of the upstream data frame consists of a preamble and a transmission payload. In the preamble, TS-A is used for frame detection and clock recovery. TS-B is the key component for the data-aided DSP, specially designed to share the same training unit for the three key DSP modules. The payload segment uses a carrier phase recovery (CPR) strategy that periodically inserts one pilot symbol into every 31 payload symbols, resulting in a 3.125% overhead.

Fig. 1(b) illustrates the receiver DSP flow based on this upstream data frame structure. When the upstream burst signals arrive, frame detection is first performed to obtain an approximate position of the frame header. After detecting the signal, a clock recovery algorithm is implemented to eliminate the sampling phase offset. Then with the aid of TS-B, FS and FOE are employed sequentially. After a matched filter, frequency domain CE is used to initialize the equalizer tap coefficients. To further recover the payload data, these tap coefficients are tracked and updated based on the LMS algorithm, embedded with pilot-base CPR. Finally, the signals are demodulated and the bit error rate (BER) is measured.

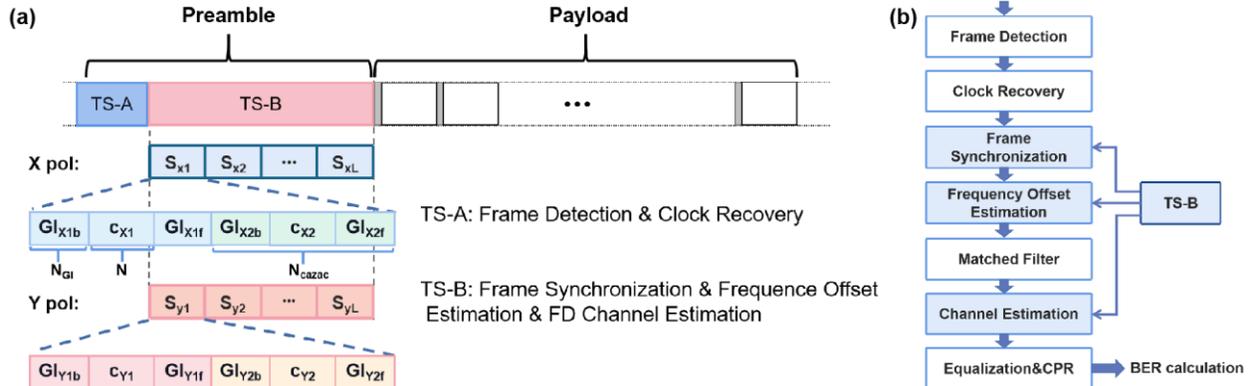

Fig. 1. (a) The proposed data frame structure. (b) DSP flow at the receiver side for burst mode detection.

In this burst reception DSP flow, frame detection can be achieved by power detection or other data-aided methods [22], [23]. Clock recovery is performed by using the Godard algorithm, which has no specific requirements for the training sequence properties or the need for accurate synchronization. Therefore, in the following parts of this section, we will mainly focus on the detailed design of TS-B and the corresponding data-aided DSP.

### A. Properties of CAZAC sequences

Since TS-B is designed mainly based on CAZAC sequences, according to [27], [28], we first define a specially simplified CAZAC sequence as

$$c(n) = \exp(j\pi n^2/N) \quad n = 1, 2, \cdots N, \tag{1}$$

where $N$ is the length of the CAZAC sequence. To facilitate the use of the Fast Fourier Transform (FFT) at the receiver, the length $N$ is defined as $2^p$ where $p$ is an integer. This CAZAC sequence has the following advantageous properties, some of which will be used in the subsequent preamble and DSP design.

#### 1) Zero Auto-correlation

The auto-correlation of the CAZAC sequence can be defined as

$$ACF_m = \sum_{n=1}^{N} c(n)c^*[(n-m) \bmod N]$$

$$= \sum_{n=1}^{m} c(n)c^*(n-m+N) + \sum_{n=m+1}^{N} c(n)c^*(n-m)$$

$$= \sum_{n=1}^{m} \exp[j\pi(2mn - m^2)/N] \exp[j\pi(2m - 2n - N)]$$

$$+ \sum_{n=m+1}^{N} \exp[j\pi(2mn - m^2)/N] \quad m = 1, 2, \cdots N, \tag{2}$$

where "$*$" represents complex conjugate and "$mod$" means modulo operation. Considering that $2m - 2n - N$ is even, (2) can be written as

$$ACF_m = \sum_{n=1}^{N} \exp(-j\pi m^2/N) \exp(j2\pi m/N)^n. \tag{3}$$

Note that $\exp(j2\pi m/N)$ is a $N$-th root of unity. According to the theorem $\sum_{n=1}^{N} r^n = 0$, where $r$ is an $N$-th root of unity and $r \neq 1$, $ACF_m$ can be simplified as

$$ACF_m = \begin{cases} N & m = N \\ 0 & m \neq N \end{cases}. \tag{4}$$

As shown in Fig. 2(a), the auto-correlation of the CAZAC sequence equals to zero for all index $m$ except $m = N$.

#### 2) Flat Spectrum

The CAZAC sequence not only has a constant amplitude in the time domain but also has a perfectly flat spectrum in the frequency domain. The FFT result $C(k)$ of the CAZAC sequence defined in (1) can be expressed as

$$C(k) = \sum_{n=1}^{N} c(n)\exp(-j2\pi kn/N)$$

$$= \sum_{n=1}^{N} \exp[j\pi(n-k)^2/N]\exp(-j\pi k^2/N)$$

$$= c^*(k)\sum_{n=1}^{N} \exp[j\pi(n-k)^2/N] \quad k = 1, 2, \cdots N. \tag{5}$$

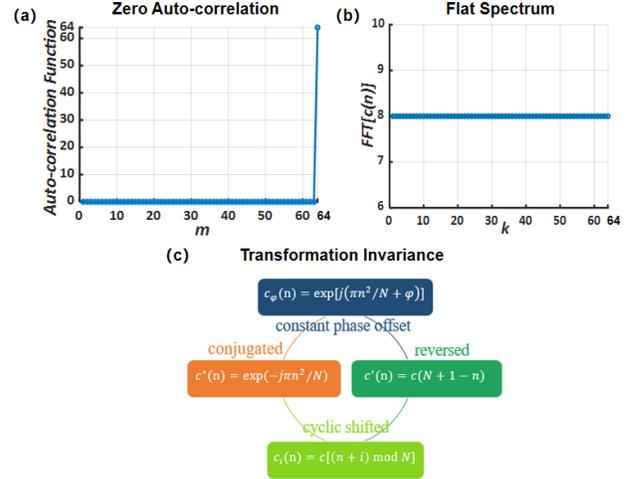

Fig. 2. Properties of proposed special CAZAC sequence.

From the definition provided in (1), the mathematical properties mentioned below can be derived:

$$c(N+n) = \exp[j\pi(N+n)^2/N] = c(n), \tag{6}$$

$$c(N-n) = \exp[j\pi(N-n)^2/N] = c(n), \tag{7}$$

$$c(n+N/2) = \exp[j\pi(n+N/2)^2/N] = \begin{cases} -c(n) & n \text{ is odd} \\ c(n) & n \text{ is even} \end{cases} \tag{8}$$

According to (6), $\sum_{n=1}^{N} \exp[j\pi(n-k)^2/N]$ can be transformed into $\sum_{n=1}^{N} c(n)$, which is a constant complex value for all indices $k$. Therefore, the FFT outcome of the CAZAC sequence can be equivalently regarded as a constant amplitude sequence. As depicted in Fig. 2 (b), the FFT result of the CAZAC sequence exhibits a flat amplitude spectrum.

#### 3) Transformation Invariance

First, we define several transformations of $c(n)$:

$$c_\varphi(n) = \exp[j(\pi n^2/N + \varphi)], \tag{9}$$

$$c_i(n) = c[(n+i) \bmod N] \quad i = 1, 2, \cdots N-1, \tag{10}$$

$$c'(n) = c(N+1-n), \tag{11}$$

where $\varphi$ is a constant phase offset. $c_i(n)$ represents the cyclic-shifted version of $c(n)$, and $c'(n)$ is the reversed version of $c(n)$. Similar to the derivation process of (2) and (5), it is straightforward to demonstrate that $c_\varphi(n)$ and $c^*(n)$ also exhibit constant amplitude, zero auto-correlation, and flat spectrum.

When $c(n)$ is transformed to $c_i(n)$, (3) can be rewritten as

$$ACF_{i,m} = \sum_{n=1}^{N} \exp[j\pi(-m^2 + 2mi)/N] \exp(j2\pi m/N)^n, \tag{12}$$

which still satisfies the zero auto-correlation property. Besides, (5) can be rewritten as

$$C_i(k) = c^*(k)\exp(j2\pi ki/N)\sum_{n=1}^{N}\exp[j\pi(n+i-k)^2/N]. \quad (13)$$

Consequently, the FFT result $C_i(k)$ maintains a flat spectrum. After we combine the mathematical property shown in (7) and the cyclic shift invariance property, it can also be deduced that the reversed version $c'(n)$ defined in (11) satisfies the inherent properties of the CAZAC sequence.

### B. TS-B Structure

As shown in Fig. 1(a), the TS-B contains $L$ repeated training units $S_{X/Y}$, each of which consists of four rearranged CAZAC blocks $[c_{X1}c_{X2}; c_{Y1}c_{Y2}]$ and their corresponding guard intervals (GI). The four training CAZAC blocks are defined as follows:

$$c_{X1} = c[1,2,\cdots N-1,N],$$
$$c_{X2} = c[N^*, N-1^*, \cdots 2^*, 1^*], \quad (14)$$

$$c_{Y1} = c[N/2+1, N/2+2, \cdots N, 1,2, \cdots N/2],$$
$$c_{Y2} = -c[N/2^*, N/2-1^*, \cdots 2^*, 1^*, N^*, \cdots N/2+1^*]. \quad (15)$$

Here $c_{X1}$ is defined as a CAZAC sequence. $c_{Y1}$ is obtained by cyclic shifting $N/2$ symbols from $c_{X1}$, $c_{X2}$ is obtained by reversing the conjugate $c_{X1}$ and $c_{Y2}$ is obtained by reversing the minus conjugate $c_{Y1}$. According to the transformation invariance property of the CAZAC sequence, these four sequences still maintain their flat spectra, which provide a distinctive advantage in overall frequency domain channel estimation. In addition, we also add a $N_{GI}$-symbol guard interval at the beginning and the end of each block. The structures of guard intervals are designed as

$$GI_{X/Y,1/2,f} = c_{X/Y,1/2}[1,\cdots,N_{GI}],$$
$$GI_{X/Y,1/2,b} = c_{X/Y,1/2}[N-N_{GI}+1, N-N_{GI}+2, \cdots N]. \quad (16)$$

The GIs serve predominantly to absorb the inter-symbol interference (ISI) between the adjacent training blocks and the interference between training blocks and payload data [24]. Therefore, an enhancement in tolerance towards CD, DGD, and bandwidth limitation can be achieved.

In this preamble, every training unit $S_{X/Y}$ can achieve FS, FOE, and CE independently. Moreover, a series of $L$ consecutive identical training units $S_{X1/Y1}, S_{X2/Y2}, \cdots, S_{XL/YL}$ are designed to further improve the DSP performance by reducing the impact of noise. The total length of this preamble is $N_{TS} = L \times N_{cazac} = L \times 2(N + 2N_{GI})$.

### C. Frame Synchronization

At the receiver side, since FOE and CE are implemented based on the known training sequences, frame synchronization should be performed first. Moreover, considering that the compensation for FO and RSOP is conducted in the subsequent stages of DSP, the FS algorithm should be designed to be tolerant to a range of FO and various SOPs.

As shown in (14) and (15), $c_{X/Y1}$ and $c_{X/Y2}$ are conjugate symmetric. The timing metric of every training unit $S_{X/Y}$ is obtained by using a sliding window to calculate the summation of product for each symmetric symbol pair. The metric can be written as

$$M_{X/Y}(n) = \left| \sum_{k=1}^{N_{cazac}} r_{X/Y}(n+k)r_{X/Y}(n+2N_{cazac}+1-k) \middle/ P_{X/Y,N} \right|, \quad (17)$$

where $N_{cazac} = N + 2N_{GI}$ and $|\cdot|$ means modulus operation. Here $r_{X/Y}(n)$ is the received signal on X/Y polarization, and $P_{X/Y,N}$ is the normalization power factor. With the design of the conjugate and symmetrical training block structure, frequency offset has no impact on the peak of $M_{X/Y}$, which has been discussed in [21] comprehensively.

When random polarization rotations are considered, new received symbols can be expressed as

$$\begin{bmatrix} r'_X \\ r'_Y \end{bmatrix} = \begin{bmatrix} cos\theta e^{j\alpha} & -sin\theta e^{j\beta} \\ sin\theta e^{-j\beta} & cos\theta e^{-j\alpha} \end{bmatrix} \begin{bmatrix} r_X \\ r_Y \end{bmatrix}, \quad (18)$$

where $\theta$ is the SOP rotation angle and $\alpha, \beta$ are random phases. According to the mathematical property shown in (8), there exists a special relation between training blocks on X/Y polarization:

$$c_{X1}(n) = \begin{cases} -c_{Y1}(n) & n \text{ is odd} \\ c_{Y1}(n) & n \text{ is even} \end{cases}, \quad (19)$$

and it can be proven that $c_{X2}$ and $c_{Y2}$ satisfy this relation as well.

According to (18) and (19), unfortunately, in certain extreme situations (e. g. $\theta = \pi/4$ and $\alpha = \beta$), inter-polarization crosstalk may offset the signal to zero at one symbol interval, causing (17) to fail. To address this problem, we calculate another two timing metrics of $r'_X + r'_Y$ and $r'_X - r'_Y$. As $r'_X$ and $r'_Y$ are not zero, $r'_X + r'_Y$ and $r'_X - r'_Y$ cannot both be zero. Therefore, at least one time function can be effective. Subsequently, we select the optimal peak among these four time functions for synchronization. This approach effectively enhances the robustness to SOP rotations.

In this preamble, a series of $L$ consecutive identical training units $S_{X1/Y1}, S_{X2/Y2}, \cdots, S_{XL/YL}$ are optionally employed. When there are multiple training units, we can multiply the timing functions of $L$ consecutive units to obtain a higher peak-to-maximum-noise ratio (PMNR).

### D. Frequency Offset Estimation

After frame synchronization, the frequency offset is estimated based on the training units $S_{X/Y}$. We first consider the situation without SOP rotations. The received symbols on each polarization can be expressed as

$$r(k) = c(k)\exp[j(2\pi f_d kT + \varphi)] + n(k), \quad (20)$$

where $c(k)$ represents the transmitted training symbols, $f_d$ is FO, $T$ is the symbol period, $\varphi$ represents the phase noise and $n(k)$ represents the additive white Gaussian noise (AWGN). Here we approximate $\varphi$ as a constant value for a short FOE training sequence. Given that $r(k)$ depends on the transmitted symbol and $c(k)c^*(k) = 1$, we multiply $r(k)$ by $c^*(k)$ to eliminate this dependence and calculate $z(k)$ as follows:

$$z(k) = r(k)c^*(k) = \exp[j(2\pi f_d kT + \varphi)][1 + \tilde{n}(k)], \quad (21)$$

where $\tilde{n}(k)$ is defined as $n(k)c^*(k)\exp[-j(2\pi f_d kT + \varphi)]$ and is statistically Gaussian noise. Subsequently, as shown in [29], we can calculate the correlations $R(m)$ as

$$R(m) = \frac{1}{N_{TS} - m} \sum_{k=m+1}^{N_{TS}} z(k)z^*(k-m)$$

$$= \exp(j2\pi m f_d T) \left[1 + \gamma(m)\right] \quad m \in [1, N_{TS}/2], \quad (22)$$

with

$$\gamma(m) = \frac{1}{N_{TS} - m} \sum_{k=m+1}^{N_{TS}} [\tilde{n}(k) + \tilde{n}^*(k-m) + \tilde{n}(k)\tilde{n}^*(k-m)]. \quad (23)$$

Here $N_{TS}$ is the total preamble length. After the averaging in (23), the random noise component $\gamma(m)$ can be effectively mitigated. Since the phase of the principal component of $R(m)$ is directly proportional to $m$, we can estimate $\overline{f_d}$ from the increment of $arg\{R(m)\}$.

$$\overline{f_d} = \frac{1}{2\pi T} \frac{1}{N_{TS}/2 - 1} \sum_{m=1}^{N_{TS}/2-1} [arg\{R(m+1)\} - arg\{R(m)\}]_{2\pi}. \quad (24)$$

However, when SOP rotations are taken into account, the received symbols should be rewritten as (18). In this case, according to (19) and (21), $z'_X(k)$ and $z'_Y(k)$ are rewritten as

$$z'_X(k) = cos\theta e^{j\alpha + j(2\pi f_d kT + \varphi_X)} - (-1)^k sin\theta e^{j\beta + j(2\pi f_d kT + \varphi_Y)},$$
$$z'_Y(k) = cos\theta e^{-j\alpha + j(2\pi f_d kT + \varphi_Y)} + (-1)^k sin\theta e^{-j\beta + j(2\pi f_d kT + \varphi_X)}. \quad (25)$$

Here we neglect statistically equivalent Gaussian noise, which could be mitigated after averaging. Similar to (22), we derive $R'_X(m)$ as

$$R'_X(m) = \frac{1}{N_{TS} - m} \sum_{k=m+1}^{N_{TS}} z'_X(k)z'^*_X(k-m)$$

$$= \frac{1}{N_{TS} - m} \sum_{k=m+1}^{N_{TS}} \exp(j2\pi m f_d T) \{cos^2\theta + (-1)^{2k-m} sin^2\theta$$

$$- sin\theta cos\theta [(-1)^k e^{j(\beta-\alpha+\varphi_Y-\varphi_X)} + (-1)^{k-m} e^{j(-\beta+\alpha-\varphi_Y+\varphi_X)}]\}. \quad (26)$$

After the summation over the index $k$, the last two terms in (26) are close to zero. $R'_Y(m)$ is calculated in the same way. Then $R'_{X/Y}(m)$ can be simplified as

$$R'_{X/Y}(m) = \begin{cases} (cos^2\theta - sin^2\theta) \exp(j2\pi m f_d T) & m \text{ is odd} \\ \exp(j2\pi m f_d T) & m \text{ is even} \end{cases}. \quad (27)$$

When the result of $cos^2\theta - sin^2\theta$ is negative and $m$ is odd, an additional phase $\pi$ will be induced in $arg\{R'_{X/Y}\}$. To ensure the feasibility across all rotation angles, we separate the odd $m$ and even $m$, selecting two-symbol increment of $arg\{R'_{X/Y}\}$ to estimate FO as

$$\overline{f_d}_{X/Y} = \frac{1}{4\pi T} \frac{1}{N_{TS}/2 - 2} \sum_{m=1}^{N_{TS}/2-2} \left[arg\{R'_{X/Y}(m+2)\} - arg\{R'_{X/Y}(m)\}\right]_{2\pi}. \quad (28)$$

In this algorithm, we estimate FO by relating $2\pi f_d(2T)$ to the increment of $arg(R'_{X/Y})$. Since the range of $arg(R'_{X/Y})$ is from $-\pi$ to $\pi$, it is worth noting that the maximum range of $\overline{f_d}$ is half of the symbol rate. To detect the positive and negative FO, we set this range to (-1/4 symbol rate, 1/4 symbol rate). In addition, as shown in (22), all the training symbols can be used to estimate FO. Therefore, increasing the number of training units $S_{X/Y}$ leads to more accurate FOE results. After calculating $\overline{f_d}_{X/Y}$ on X/Y polarizations, we finally average them as the final result of FOE.

## E. Channel Estimation

CE aims to estimate linear channel effects and then accelerate the equalization convergence process. In this section, we first utilize the training blocks to estimate the channel transfer function $H(\omega)$. Subsequently, we calculate the initial FD equalizer coefficients $W(\omega)$ from $H(\omega)$ using the zero-forcing (ZF) criterion. In this way, the overall frequency-domain CE can effectively achieve the pre-convergence of equalization.

As mentioned earlier, the CAZAC sequence with its unique flat spectrum provides a distinct advantage in comprehensive channel estimation over the whole signal bandwidth. In addition to RSOP, the FD channel estimation encompasses many other linear channel effects, including bandwidth limitation, CD, PMD, and PDL. The overall impact of these linear effects can be described as a multiplication of channel transfer functions in the FD:

$$H(\omega) = H_{Rx}(\omega)H_{CD}(\omega)H_{PMD}(\omega)H_{PDL}(\omega)H_{RSOP}(\omega)H_{Tx}(\omega), \quad (29)$$

where $H_{Tx/Rx}(\omega)$ represents the transfer functions of the pulse shaper at the transmitter and the receiver, and $H_{RSOP}(\omega)$ represents the random SOP rotation after fiber transmission.

The CD transfer function is defined as

$$H_{CD}(\omega) = \exp\left(-j\frac{D\lambda^2\omega^2}{4\pi c}z\right), \quad (30)$$

where $D$ is the dispersion parameter, $\lambda$ is the central wavelength, $z$ is the link length, and $c$ is the speed of light.

The DGD is simulated by the first-order PMD model:

$$H_{PMD}(\omega) = \begin{bmatrix} cos\alpha & -sin\alpha \\ sin\alpha & cos\alpha \end{bmatrix}$$

$$\times \begin{bmatrix} \exp(j\omega\Delta\tau/2) & 0 \\ 0 & \exp(-j\omega\Delta\tau/2) \end{bmatrix}$$

$$\times \begin{bmatrix} cos\alpha & -sin\alpha \\ sin\alpha & cos\alpha \end{bmatrix}^{-1}, \quad (31)$$

where $\Delta\tau$ represents the DGD between the fast and slow axes.

The PDL can be modeled in a similar way [30]:

$$H_{PDL}(\omega) = \begin{bmatrix} cos\beta & -sin\beta \\ sin\beta & cos\beta \end{bmatrix} \begin{bmatrix} \sqrt{1+\gamma} & 0 \\ 0 & \sqrt{1+\gamma} \end{bmatrix} \begin{bmatrix} cos\beta & -sin\beta \\ sin\beta & cos\beta \end{bmatrix}^{-1}. \quad (32)$$

Here $\gamma$ is the PDL parameter, which is used in the PDL definition:

$$PDL = 10\log_{10}\left(\frac{1+\gamma}{1-\gamma}\right). \quad (33)$$

The overall linear channel transfer function $H(\omega)$ can be obtained based on each training unit $S_{X/Y}$. In a noiseless condition, the received TS can be modeled as

$$\begin{bmatrix} R_{X1}(\omega_k) & R_{X2}(\omega_k) \\ R_{Y1}(\omega_k) & R_{Y2}(\omega_k) \end{bmatrix} = \begin{bmatrix} H_{xx}(\omega_k) & H_{xy}(\omega_k) \\ H_{yx}(\omega_k) & H_{yy}(\omega_k) \end{bmatrix} \times \begin{bmatrix} C_{X1}(\omega_k) & C_{X2}(\omega_k) \\ C_{Y1}(\omega_k) & C_{Y2}(\omega_k) \end{bmatrix}. \quad (34)$$

Here $C(\omega_k)$ represents the $k$-th spectrum element of the two-fold upsampled CAZAC training blocks, which is obtained by adding a zero between each of the symbol of $[c_{X1} \, c_{X2}; c_{Y1} \, c_{Y2}]$. $R(\omega_k)$ represents the $k$-th spectrum element of corresponding received samples. Here we define $k = 0,1,\cdots, M-1$, where $M$ is the length of the upsampled CAZAC training block. Subsequently, the channel transfer matrix can be calculated by

$$\begin{bmatrix} H_{xx}(\omega_k) & H_{xy}(\omega_k) \\ H_{yx}(\omega_k) & H_{yy}(\omega_k) \end{bmatrix} = \begin{bmatrix} R_{X1}(\omega_k) & R_{X2}(\omega_k) \\ R_{Y1}(\omega_k) & R_{Y2}(\omega_k) \end{bmatrix} \times \begin{bmatrix} C_{X1}(\omega_k) & C_{X2}(\omega_k) \\ C_{Y1}(\omega_k) & C_{Y2}(\omega_k) \end{bmatrix}^{-1}, (35)$$

where $[\cdot]^{-1}$ represents matrix inversion operation. Subsequently, we use the ZF criterion to obtain the initial coefficients $W(\omega_k)$ of the equalizer. According to [24], the normalized $W(\omega_k)$ can be calculated by

$$\begin{bmatrix} W_{xx}(\omega_k) & W_{xy}(\omega_k) \\ W_{yx}(\omega_k) & W_{yy}(\omega_k) \end{bmatrix} = \left[ \frac{1}{|H_{Rx}(\omega_k)|^2} \begin{bmatrix} H_{xx}(\omega_k) & H_{xy}(\omega_k) \\ H_{yx}(\omega_k) & H_{yy}(\omega_k) \end{bmatrix} \right]^{-1}, (36)$$

where $|\cdot|$ means modulus operation. $H_{Rx}(\omega)$ is the transfer function at the receiver, contributing to normalize $H(\omega)$. Even though CE can be independently achieved based on a single training unit $S_{X/Y}$, in noisy scenarios, it is preferable to average $W_L(\omega)$ over $L$ CE processes to mitigate the impact of random noise.

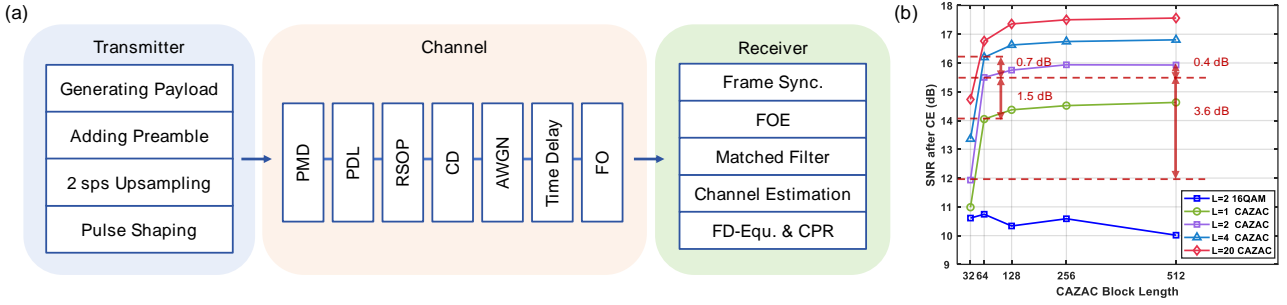

Fig. 3. (a) Simulation setup. (b) SNR after CE versus the CAZAC block length for different unit numbers.

### F. Frequency-domain Equalization and Pilot-aided CPR

FD equalization is executed to recover subsequent payload data. The block length of the equalizer equals to the length of each training block CAZAC $N$ and the tap length equals to twice of the block length (50% overlap-save). Before processing the payload data, the proposed preamble can also be employed repetitively as a training sequence. The coefficients are iteratively updated based on the LMS algorithm to facilitate full convergence. Consequently, when the system starts processing the payload data, the equalizer can seamlessly switch to the decision-directed mode and continue tracking the coefficients using the LMS algorithm.

Before calculating the error function of the LMS algorithm, the carrier phase noise is compensated by using a pilot-based maximum likelihood (ML) CPR algorithm. Here we first utilize the pilot symbol between every 31 payload symbols to estimate a coarse phase noise $\varphi_{coarse}$. After this coarse compensation, the residual phase noise is estimated using the ML algorithm. This CPR algorithm effectively removes carrier phase noise to ensure proper tracking of the equalizer coefficients.

## III. SIMULATION RESULTS AND DISCUSSIONS

As mentioned above, every training unit $S_{X/Y}$ can perform FS, FOE, and CE independently, while increasing the number of training units can further improve the DSP performance. Consequently, in this section, we first evaluate the CE performance with different CAZAC block lengths and training unit numbers to determine an optimal preamble configuration. Additionally, the feasibility and robustness of FS, FOE, and CE are also verified. The overall simulation setup is shown in Fig. 3(a). Several optical linear effects including PMD, PDL, RSOP, and CD are added in this system. In addition, random time delay, FO, and additive white Gaussian noise are also present.

TABLE I
SIMULATION PARAMETERS OF FIG. 3(b)

| Simulation parameters | Values |
|---|---|
| modulation format | 16QAM |
| symbol rate | 15 Gbaud |
| payload length | $2^{15}$ |
| RRC roll-off factor | 0.1 |
| DGD $\Delta\tau$ | 30 ps |
| PDL | 3 dB |
| RSOP angle $\theta$ | random |
| CD | 340 ps/nm |
| AWGN | 18 dB |
| time delay | 0 |
| FO | 0 |

To select the optimal preamble configuration, we evaluate the CE performance versus the CAZAC block length for different unit numbers. Fig. 3(b) exhibits the average signal-to-noise ratio (SNR) of the payload signals over 100 trials, which are only compensated by the initial equalization coefficients from CE without any subsequent update. As the length of the proposed CAZAC sequence is defined as $2^p$ symbols, the tested CAZAC block lengths are set to [32, 64, 128, 256, 512], and the unit number $L$ is set to [1, 2, 4, 20]. The guard interval length is fixed at 2 symbols. In this system, the FS and FOE modules are omitted, while the remaining modules are identical to those in Fig. 3(a). The detailed simulation parameters are summarized in Table I. In addition, to verify the superiority of the CAZAC sequence, the CE performances based on ordinary 16QAM training blocks are also presented in Fig. 3(b). It can be seen that CE based on CAZAC blocks exhibits significantly superior performances. Consequently, the scheme based on the CAZAC sequence is adopted in this design.

As shown in Fig. 3(b), with the increase of the training unit number $L$, signals recovered by CE exhibit a higher SNR. This gain mainly comes from the mitigation of random noise impact via averaging the CE coefficients. Compared to CE based on only one training unit, averaging two units induces considerable SNR improvement. However, the performance of CE will gradually saturate with a further increased number of units. For example, when the CAZAC block length is set to 64, the CE with two training units shows an SNR gain of 1.5 dB compared to the CE with only one unit. However, this gain diminishes to 0.7 dB when the number of units further increases to four. Moreover, recovered signals exhibit a higher SNR with a longer CAZAC block length. Likewise, the SNR also reaches saturation with the increase of block length. As depicted in Fig. 3(b), the preambles with two training units exhibit a 3.6 dB SNR gain when the block length increases from 32 to 64, while only a 0.4 dB gain when increasing from 64 to 512. Since these initialized coefficients can be subsequently updated to achieve the best SNR performance, a scheme with two training units and 64-symbol CAZAC blocks is preferred to make a trade-off between the preamble length and CE accuracy.

As averaging two training units can provide significant gain in CE, we set the unit number to 2 for subsequent simulations, where we further explore the performance of FS and FOE versus the CAZAC block length. When we evaluate the performance of FS and FOE, only random RSOP and 18 dB AWGN are added in the simulation. The robustness of FS and FOE against other optical linear effects such as CD, DGD, and PDL will be verified subsequently in the result shown in Fig. 7.

For frame synchronization, as shown in Fig. 4(a), PMNR is used to measure the quality of synchronization peaks. We first test the synchronization performance across various CAZAC block lengths. As indicated in Fig. 4(b), when the block length is 64, the average timing metric over 50 trials exhibits a high-quality peak with an over 10 dB PMNR, which can ensure precise synchronization. In Fig. 4(c) and Fig. 4(d), we set the CAZAC block length to 64 and test the synchronization performance in different frequency offset and RSOP conditions. As shown in Fig. 4(c), owing to the symmetric conjugate training blocks, the PMNR exhibits independence of frequency offset. For 15 Gbaud signals, we confirm that the average PMNR can remain over 10 dB within a frequency offset range of -3 GHz to 3 GHz. Moreover, as mentioned in Section 2, we calculate another two timing metrics of $r'_X + r'_Y$ and $r'_X - r'_Y$ to ensure the proper synchronization in all RSOP situations. Fig. 4(d) shows the performance of the four timing metrics: $r'_X$, $r'_Y$, $r'_X + r'_Y$ and $r'_X - r'_Y$. Here we select the timing metric with the highest PMNR as the optimal metric used for synchronization. This optimal timing metric exhibits an over 7.2 dB PMNR when the RSOP angle θ increases from 0 to π, verifying the RSOP tolerance.

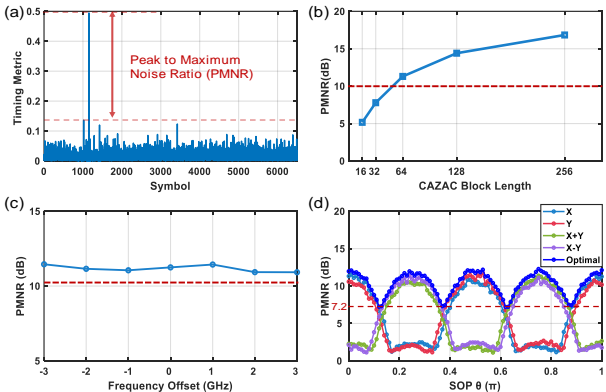

Fig. 4. (a) Result of the timing metric. (b) PMNR of the timing metric versus CAZAC block length (L=2). (c) PMNR versus frequency offset (L=2, Block length=64). (d) PMNR versus RSOP angle θ (L=2, Block length=64).

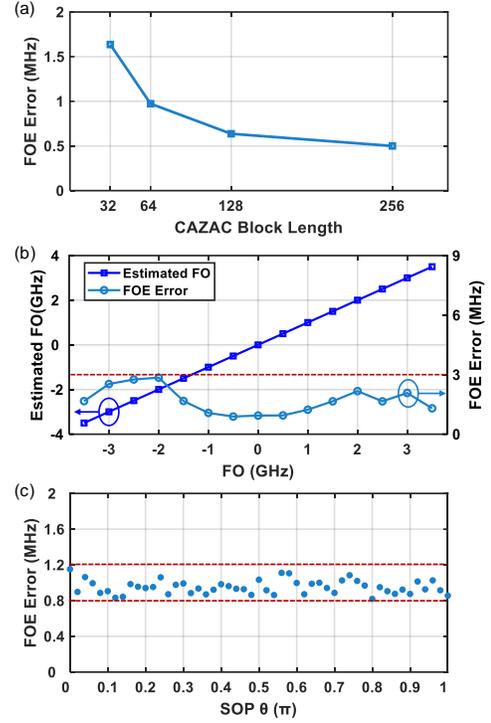

Fig. 5. Absolute value of FOE error versus (a) CAZAC block length (L=2, FO=200 MHz), (b) FO (L=2, Block length=64), (c) RSOP angle θ (L=2, Block length=64, FO=200 MHz).

The performance of FOE is depicted in Fig. 5. Fig. 5(a) provides the average absolute value of FOE error over 100 trials versus the CAZAC block length, in the presence of a 200 MHz frequency offset. As shown in Fig. 5(a), with a CAZAC block length of 64, the average FOE error is approximately 1 MHz, which is sufficiently low to ensure proper channel estimation and can be effectively compensated by subsequent CPR process. Fig. 5(b) depicts the average FOE error versus frequency offset. As mentioned in Section II, the FOE range is from -1/4 symbol rate to 1/4 symbol rate, so here we set this range to (-3.5 GHz, 3.5 GHz) for the 15 Gbaud signal. As shown in Fig. 5(b), the average FOE error remains below 3 MHz across such a large estimation range. In Fig. 5(c), the robustness against SOP rotations is validated. As the RSOP angle θ incrementally increases by 0.02π from 0 to π, the average FOE error remains stable, with a maximum fluctuation of 0.4 MHz.

For channel estimation, we first verify the feasibility of estimating RSOP, which is a prevalent and critical linear effect in short-distance scenarios. Fig. 6 shows the root mean square error (RMSE) of the output payload signals after equalization with and without CE initialization. The simulation setup is the same as the one shown in Fig. 3(a), where the RSOP angle is set to 0.25π and other linear effects are omitted. Random time delay, frequency offset (200 MHz), and AWGN (SNR=18 dB) are also added in this system. In the subsequent simulations, the CAZAC block length is set to 64 symbols, the GI length is set to 2 symbols, and the unit number is set to 2. The equalizer is configured with a tap number of 128 and a block length of 64, with a 50% overlap. For the equalizer with CE, the initial equalizer coefficients are first obtained from the CE module. To improve the pre-convergence performance, the preamble is also utilized as a training sequence, enabling further updates of the coefficients through the LMS algorithm. After the initialization, the equalizer operates in decision-directed mode throughout the payload data processing. For the standard equalizer without CE, 100 known training blocks are used to converge the equalizer coefficients, after which the equalizer switches to decision-directed mode. In Fig. 6, "Block Index" refers to the index of the output payload signal blocks, with each block containing 32 symbols.

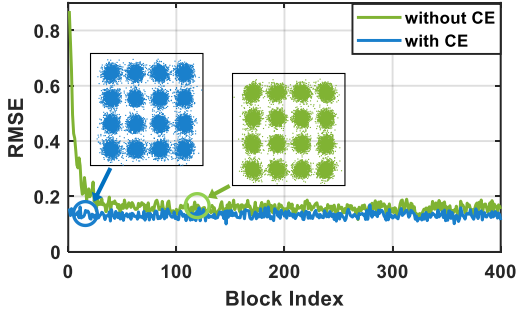

Fig. 6. Root mean square error versus block index after equalization (RSOP angle $\theta = 0.25\pi$).

In Fig. 6, it can be seen that the equalizer with CE has already converged at the beginning while the equalizer without CE requires nearly 100 blocks to converge. More CE performances under other conditions are shown in Fig. 7. Here we select the RMSE of the first 32 symbols as a metric to measure the pre-convergence performance.

In Fig. 7(a-d), only one specific optical effect exists in every subplot, while others are omitted. Here the RSOP angle $\theta$ defined in (18) varies from 0 to $\pi$, and the DGD time delay $\Delta\tau$ defined in (31) is set from 10 ps to 90 ps. In Fig. 7(a-c), it can be seen that the RMSE in the system with CE stably maintains a relatively small value, exhibiting superior pre-convergence performance under severe conditions. With the aid of CE based on the 272-symbol preamble, this 15 Gbaud system can tolerate CD up to 1360 ps/nm, DGD up to 80 ps, and all RSOP scenarios. In Fig. 7(d), as the PDL defined in (33) increases from 1 dB to 7 dB, the RMSE in the system with CE increases much more slowly compared to the system without CE, confirming that CE effectively mitigates the impact of PDL. It is worth noting that since FS and FOE are performed before CE, the feasibility of these three DSP modules can be verified simultaneously in Fig. 7.

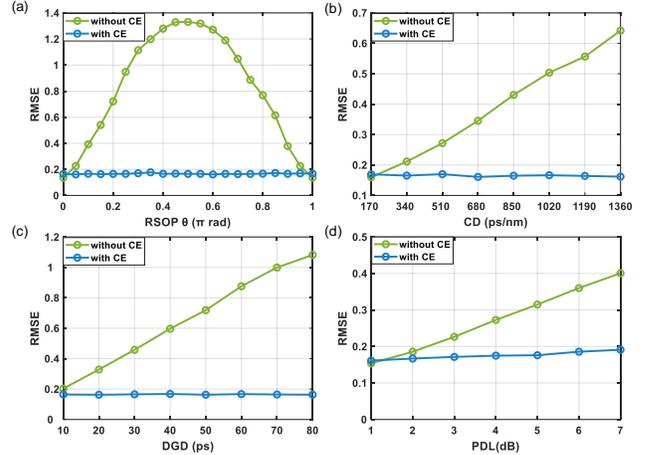

Fig. 7. RMSE in the first block under different (a) RSOP, (b) CD, (c) PMD, and (d) PDL conditions.

## IV. EXPERIMENT RESULTS AND DISCUSSIONS

### A. Experimental Setup

To validate the feasibility of this scheme, a 15 Gbaud DP-16QAM coherent system with the proposed preamble and corresponding DSP was established. Fig. 8 shows the experimental setup. At the transmitter (Tx) side, a burst data frame comprising a 272-symbol preamble and a $2^{15}$-symbol 16QAM payload was initially generated. In this experiment, the CAZAC block length $N$ was set to 64, the guard interval length was set to 2, and the training unit number $L$ was chosen as 2. The total preamble length amounted to 272 symbols. Subsequently, the data frame underwent pulse shaping using a root-raised cosine (RRC) filter with a 0.1 roll-off factor. Following pulse shaping, samples were resampled to align with the sampling rate of the arbitrary waveform generator (AWG). After Tx DSP, electrical analog signals were generated using a 120 GSa/s AWG. These electrical signals were then fed into a DP in-phase and quadrature (IQ) modulator and converted to optical signals. An external cavity laser (ECL) is adopted with a central wavelength of 1550 nm and a nominal linewidth of less than 100 kHz. After being amplified to 0 dBm by an erbium-doped fiber amplifier (EDFA), the optical signal entered an 11 km standard single-mode fiber (SSMF).

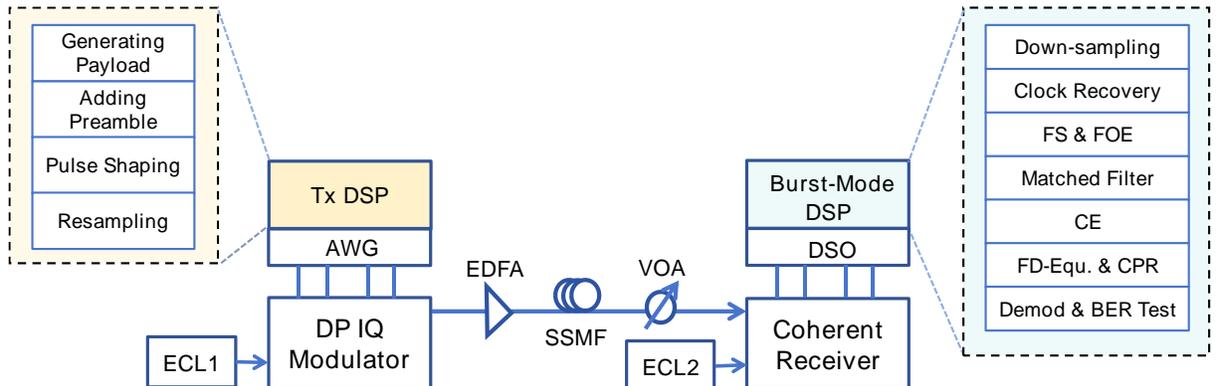

Fig.8. Experimental setup of a 15 Gbaud DP-16QAM coherent system with the proposed preamble and corresponding Rx-DSP.

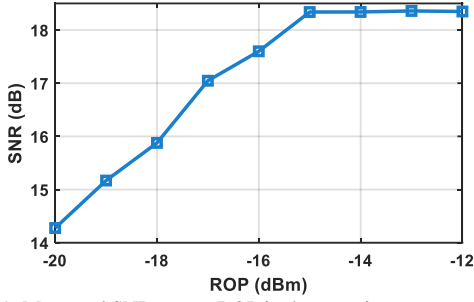

Fig. 9. Measured SNR versus ROP in the experiment.

At the receiver side, the optical signal was first attenuated by a variable optical attenuator (VOA) to adjust the received optical power (ROP). Following coherent detection utilizing another similar ECL, a real-time digital storage oscilloscope (DSO) with a sampling rate of 100 GSa/s was employed to digitize the waveform. Finally, the output signal was processed offline based on the burst-mode DSP.

### B. Experimental Results and Discussions

The experimental results are shown in Fig. 9-12. Fig. 9 provides the overall SNR of the payload signals versus ROP in the experiment. Fig. 10 presents the FOE results with a -20 dBm ROP. In the experiment, an approximately 200 MHz frequency offset was induced by devices. Therefore, to estimate the actual system FO, a classical blind 4th power FOE algorithm was employed. As shown in Fig. 10, our FOE algorithm exhibits accurate estimations with respect to the reference 4th power FOE algorithm. The average error is within 0.4 MHz in a range from -1.5 GHz to 1.5 GHz.

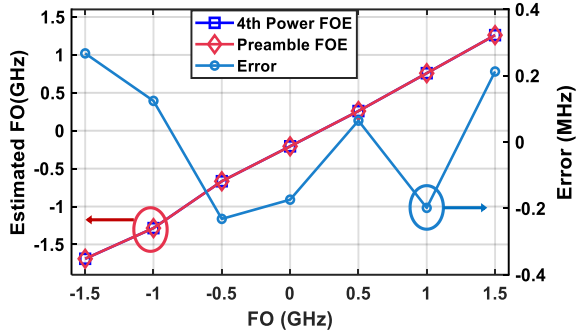

Fig. 10. Preamble-based FOE and 4th power FOE results versus FO in the experiment.

Fig. 11 shows the performance of the FD-equalizer with a -20 dBm ROP. In the equalizer, the tap number and the block length were 128 and 64 (50% overlap). Each output block contained 32 symbols. For the equalizer with CE, we first obtained the initialized equalizer coefficients from the CE module based on the proposed preamble. To improve pre-convergence performance, the same preamble was reused as a training sequence, enabling further coefficient updates via the LMS algorithm. After this initialization process, the decision-directed mode was employed for the payload data. For the normal equalizer, 100 known training blocks were first utilized to converge the equalizer coefficients, after which the decision-directed mode was conducted.

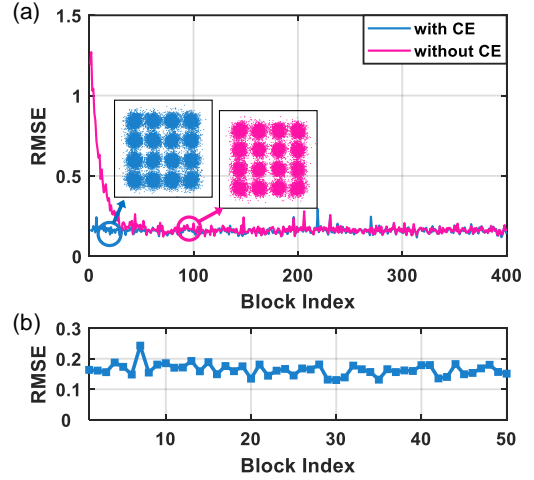

Fig. 11. (a) RMSE versus block index. (b) Details of RMSE in the first 50 blocks.

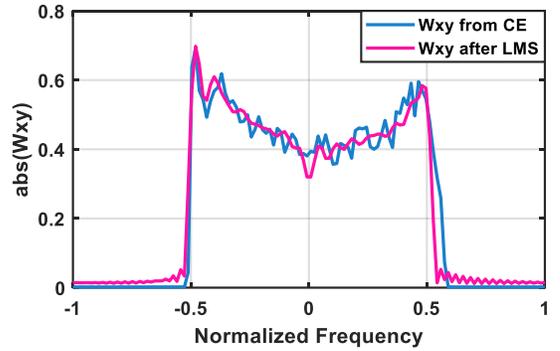

Fig. 12. Absolute value of $W_{xy}$ calculated from CE (blue line) and obtained after equalization convergence without CE (pink line).

As illustrated in Fig. 11(a), the equalizer with CE can greatly accelerate the convergence process. Fig. 11(b) shows the RMSE with CE in the first 50 blocks. It can be seen that the equalizer with CE has already converged at the beginning whereas the equalizer without CE requires nearly 50 blocks to converge. In addition, Fig. 12 shows the absolute values of the initialized coefficients calculated from CE and the coefficients obtained after equalization convergence without CE. While the initialized coefficients show some fluctuations due to random noise, they closely align with the fully converged coefficients, thereby verifying the feasibility and validity of the CE module.

## V. CONCLUSION

In this work, we proposed a novel, efficient preamble design based on the CAZAC sequence and corresponding DSP for the upstream burst-mode detection in coherent PONs. This preamble facilitates comprehensive frequency-domain channel estimation, enabling more accurate equalizer initialization and thereby significantly reducing the required preamble length for equalization convergence. Besides, the preamble efficiently integrates frame synchronization, frequency offset estimation, and channel estimation by utilizing shared training units, further minimizing the required preamble length. In this paper, the structure of the proposed preamble and the principles of the corresponding DSP are introduced in detail. We also verify the feasibility and robustness of the DSP under various conditions, including RSOP, CD, PMD, and PDL. As a proof-of-concept, we built a 15 Gbaud DP-16QAM coherent system with the proposed 272-symbol preamble and validated its superior convergence acceleration. In conclusion, this proposed preamble provides a promising solution to the challenge of upstream burst-mode detection in next-generation coherent PONs.